\documentstyle[aps,prc,preprint]{revtex}

\begin{document}
\draft
\tightenlines  
\title{$\nu d\rightarrow \mu^- \Delta^{++}n$ Reaction and Axial Vector
$N-\Delta $ Coupling }  
\author{L. Alvarez-Ruso, S.K. Singh\cite{singhper} 
and M.J. Vicente Vacas}
\address{Departamento de F\'{\i}sica Te\'{o}rica and IFIC, Centro Mixto 
Universidad de Valencia-CSIC,\\ 46100 Burjassot, Valencia, Spain}
\date{\today}
\maketitle

\begin{abstract} The reaction $\nu d \rightarrow \mu^- \Delta^{++} n$
is studied in the region of low $q^2$ to investigate the effect of
deuteron structure and width of the $\Delta$ resonance on the
differential cross section. The results are used to extract the axial
vector $N-\Delta$ coupling $C^{A}_5$ from the experimental data on 
this reaction. The possibility to determine this coupling from electroweak 
interaction experiments with high intensity electron accelerators is 
discussed.  
\end{abstract}


\section{Introduction}
The study of electromagnetic and weak couplings in the $N-\Delta$ transition
amplitude can provide valuable information about the hadron structure.
For example, the electromagnetic couplings in the magnetic dipole $(M1)$
and electric quadrupole $(E2)$ transition amplitudes, determined from the
experiments on photo and electroproduction of the $\Delta$ resonance are
found to be about $30\%$ larger than those computed in many theoretical
models of hadron structure \cite{1}. To explain this discrepancy is a
challenging task for these models. A similar comparison between theoretical
and experimental values of the various couplings in the weak transition
amplitude has not been made, even though there exists considerable literature
on the study of weak $N-\Delta$ transitions \cite{2}. However, in a recent
paper, Hemmert, Holstein and Mukhopadhyay(HHM) \cite{3}, using the low $q^2$
data from the Argonne National Laboratory (ANL) experiment of Barish
{\it et al.} \cite{4} and the Brookhaven National Laboratory (BNL) experiment
of Kitagaki {\it et al.} \cite{5} on the reaction
$ \nu d \rightarrow \mu^- \Delta^{++} n$, have determined the value of
the axial vector $N-\Delta$ coupling $C^{A}_5$. They find that, in the weak
sector too, the experimental value of $C^{A}_5$ is about $30\%$ larger than
the theoretical estimates obtained in most of the quark models. This value
is, however, consistent with the value obtained in a calculation that uses the
hypothesis of partial conservation of axial current (PCAC), when the
experimental value is used for the $g_{\Delta N \pi}$ coupling.
The underestimation of the electromagnetic and weak couplings in the
$N-\Delta$ transitions may be a manifestation of the large violations of 
$SU(6)$ symmetry, while maintaining the chiral symmetry of the Lagrangian, and 
needs further investigation. On the experimental side, a better determination 
of these couplings might become available in near future, when the 
electromagnetic and weak interaction reactions planned to be studied at high 
intensity electron accelerators are performed \cite{6}.

In this paper, we undertake the determination of $C^{A}_5$ using the data
from the BNL experiment of Kitagaki {\it et al.} \cite{5} on the ratio of the 
differential cross sections for the inelastic 
$\nu d \rightarrow \mu^- \Delta^{++}n$ and the quasielastic 
$\nu d \rightarrow \mu^- p p$  reactions. We also analyze the experimental
results from the ANL experiment of Radecky {\it et al.} \cite{7}, which has 
about three times more events than the experiment of Barish {\it et al.} 
\cite{4}. In the inelastic reaction, all the experimental analyses 
\cite{4,5,7} exclude the region of very low $|q^2|$ i.e. 
$|q^2| \leq 0.1$ GeV$^2$. In this region, the nuclear
corrections due to the deuteron target have not been calculated. We take into 
account the effect of deuteron structure in the present work. We also study 
the effect of the width of the $\Delta $ resonance on the differential cross 
section, and its influence on the determination of $C^{A}_5$ using an energy 
dependent P-wave width for the $\Delta$. In the earlier analyses of this 
reaction \cite{4,5,7}, an energy dependent S-wave width was used. These 
effects were not included in the analysis of HHM \cite{3}, which could 
influence the determination of $C^{A}_5$, specially when the low $q^2$ data is 
used for the ratio of the differential cross section of the
inelastic reaction $\nu d \rightarrow \mu^- \Delta^{++} n$ and  the 
quasielastic reaction $\nu d \rightarrow \mu^- p p $. The analysis presented 
here brings out in detail the various uncertainties involved in the 
extraction of $C^{A}_5$ from the data, when extrapolated to $q^2=0$.

In Sec. II, we calculate the effects of deuteron structure and width of the
$\Delta$ resonance on the differential cross sections. We determine the 
value of $C^{A}_5$ in Sec. III, where the possibility of extracting
it from electron scattering experiments is also discussed. The Sec. IV 
provides a summary of the results presented in this paper.

\section{Differential cross section}
\subsection{Differential Cross Section for 
$\nu p \rightarrow \mu^- \Delta^{++}$}

 The weak $N-\Delta$ transition is described in terms of eight form factors
$C^{V,A}_i(i=3-6)$, where superscripts $V$ and $A$ refer to the vector and 
axial vector form factors respectively. In the standard notation 
\cite{8,9a,9}, the amplitude ${\cal M}$ is written as  
 
\begin{equation} 
{\cal M}= {{G}\over{\sqrt{2}}}\, \cos \theta_{c}\, l_{\alpha} J^{\alpha} \,, 
\end{equation} 
with 

\begin{equation} 
l_{\alpha}= \bar{u}(k') \gamma_{\alpha} (1 -\gamma_{5}) u(k)\,, 
\end{equation} 
and 

\begin{eqnarray}
 \nonumber 
 J^{\alpha}=& \sqrt{3} \bar{\psi_{\mu}}(p') \{ [
  {{C_3^V}\over{M}} (g^{\mu \alpha} q \!\!\! / - q^{\mu} \gamma^{\alpha})+
  {{C_4^V}\over{M^2}} (g^{\mu \alpha} q\cdot p' - q^{\mu} p'^{\alpha}) 
  \\ \nonumber 
  &+ {{C_5^V}\over{M^2}} (g^{\mu \alpha} q\cdot p - q^{\mu} p^{\alpha})
  ] \gamma_{5}\\ 
  &+ {{C_3^A}\over{M}} (g^{\mu \alpha}q \!\!\! / - q^{\mu} \gamma^{\alpha})+
 {{C_4^A}\over{M^2}} (g^{\mu \alpha} q\cdot p' - q^{\mu} p'^{\alpha})+
 {C_5^A} g^{\mu \alpha}\\ 
 \nonumber 
  &+ {{C_6^A}\over{M^2}} q^{\mu} q^{\alpha} \} u(p)\,,
\end{eqnarray}  
where $M$ is the nucleon mass; $\psi_{\mu}(p')$ and $u(p)$ are the 
Rarita Schwinger and Dirac spinors for $\Delta$ and nucleon of momentum $p'$ 
and $p$; $q=p'-p=k-k'$ is the momentum transfer.
The weak form factors $C^{V}_i(i=3-6)$ are obtained using the conserved vector 
current (CVC) hypothesis, which requires $C^{V}_6=0$ and relates the remaining 
three form factors to the various amplitudes in the photo and electroproduction
of the $\Delta $ resonance. From the experimental data on these processes,
the following values of the vector form factors are obtained, which are used
in the analysis of the neutrino scattering experiments \cite{4,5,7,9}:
 
\begin{equation} 
C_5^V=0,\;\;\; C_4^V=-{{M}\over{M'}}C_3^V\,,
\end{equation}
with
\begin{equation}
C_3^V(q^2) = {{2.05}\over{(1-q^2/0.54{\rm GeV}^2)^2}}.
\end{equation}
Here $M'$ is the mass of $\Delta$ resonance.
The weak axial form factors $C^{A}_i(i=3-5)$ are determined by fitting the
available data on the differential cross section $d\sigma/ dq^2$
in neutrino scattering, mainly from the  deuteron target, in order to minimize 
the nuclear corrections. However, these values of the form factors are also
compatible with the data on neutrino scattering from nuclear targets\cite{10}. 
It is to be noted that $C^{A}_6$ is not determined from these experiments as 
it is proportional to the lepton mass, which is neglected in these analyses. 
Instead, this form factor is determined in terms of $C^{A}_5$ using the 
hypothesis of  PCAC. The values of the axial form factors most often used in 
the analysis of the neutrino experiments are \cite{4,5,6,7,9a,9,10}

\begin{equation}
C_{i=3,4,5}^A(q^2) = {{C_i^A(0)\left[ 1-{{a_i q^2}\over{b_i-q^2}} \right] }
{\left( 1- {{q^2}\over{M_A^2}}\right)^{-2}}} ,
\end{equation}
and
\begin{equation}
C_6^A(q^2) = C_5^A {{M^2}\over{m_{\pi}^2-q^2}}\,,
\end{equation}
with $C_3^A(0)=0$, $C_4^A(0)=-0.3$, $C_5^A(0)=1.2$, $a_4=a_5=-1.21$,
$b_4=b_5=2$ GeV$^2$ and $M_A$ is treated as a free parameter. For our present
purpose, we take $M_A=1.28$ GeV \cite{5}.
Using the matrix element of Eqs. (1-3), the differential cross section
is written as 

\begin{equation}
\frac{d^2 \sigma}{d q^2 d k'^0} = \frac{1}{128 \pi^2} \frac{M}{M'} 
\frac{1}{(s-M^2)^2} G^2 \cos^2\theta_c L_{\alpha \beta} J^{\alpha \beta}
\frac{\Gamma(W)}{(W-M')^2+\Gamma^2(W)/4} \,, 
\end{equation}
with
\begin {equation}
L_{\alpha \beta} = k_\alpha k'_\beta + k_\alpha' k_\beta
- g_{\alpha \beta} k . k' + i \epsilon_{\alpha \beta
\gamma \delta} k^\gamma k'^{\delta}\,,
\end {equation}
and
\begin {equation}
J_{\alpha \beta} = \bar{\Sigma} \Sigma J_\alpha^\dagger J_\beta\,,
\end {equation}
where the summation is performed over the hadronic spins, using a spin $3/2$
projection operator $P_{\mu \nu}$ given by

\begin {equation}
P_{\mu \nu} = -\frac{p' \!\!\!\! / + M'}{2 M'} \left( g_{\mu \nu}-
\frac{2}{3}\frac{p'_{\mu}p'_{\nu}}{M'^2} +
\frac{1}{3}\frac{p'_{\mu}\gamma_{\nu} - p'_{\nu}\gamma_{\mu}}{M'}
-\frac{1}{3}\gamma_{\mu}\gamma_{\nu} \right).
\end{equation}
In Eq. (8), $s=(p+k)^2$, $W$ is the $\Delta$ invariant mass $W^2=p'^2$  
and $\Gamma(W)$ its decay width given by \cite{33}

\begin{equation}
\Gamma= \Gamma_0 \frac{M'}{W} \frac{q^3_{cm}(W)}{q^3_{cm}(M')}\,,
\end{equation}
with $\Gamma_0 = 120$ MeV \cite{15a} and $q_{cm}(W)$ the
modulus of the pion momentum in the rest frame of a $\Delta$ with invariant
mass $W$; $k'_0$ is the muon energy in the laboratory frame.

\subsection{Effect of Deuteron Structure}

When the reaction takes place in deuteron, i.e.
$\nu(k) + d(p) \rightarrow \mu^-(k') + \Delta^{++}(p'_1) + n(p'_2)$,
the differential cross section in the impulse approximation is calculated to 
be 

\begin{equation}
\frac{d^2 \sigma}{d q^2 d k'^0} = \frac{1}{128 \pi^2} \frac{M_d^2}{M' 
(s-M_d^2)^2} G^2 \cos^2\theta_c L_{\alpha \beta} J^{\alpha \beta} 
\int \frac{d \bbox{p'_2}}{(2 \pi)^3 p'^0_{2}} 
\frac{\Gamma(W)}{(W-M')^2+\Gamma^2(W)/4} \phi^2(|\bbox{p'_2}|) \,,
\end{equation}
where $M_d$ is the deuteron mass and $\phi(|\bbox{p'_2}|)$ is the Fourier
transform of the deuteron radial wave function. This expression is derived
assuming the neutron to be spectator, and neglecting meson exchange currents
and final state interactions. The contribution of these effects on the
differential cross section $d \sigma / d q^2$ has been studied earlier 
for the case of the quasielastic reaction\cite{14} and found to be small in the
kinematical region considered here. Using Eq. (13), we calculate the 
differential cross section for the reaction 
$\nu d \rightarrow \mu^- \Delta^{++}n$ for various deuteron wave functions 
corresponding to Hulthen \cite{11}, Paris \cite{12} and Bonn
\cite{13} potentials, and compare them with the differential cross section
results for the free case, calculated from Eq. (8). 

The results for $d \sigma/d q^2$ as a function of $Q^2=-q^2$ for the
incident neutrino energy $E_{\nu}=1.6$ GeV are shown in  Fig. 1.
We see that the deuteron effects are small, not exceeding $8\%$ even at
low $Q^2$ i.e. $Q^2 < 0.1$ GeV$^2$. This is the region where 
they give a large reduction in the quasielastic
reaction $\nu d \rightarrow \mu^- p p$ \cite{14}. The different 
behaviour of deuteron effects in these two reactions is due to the 
nature  of the vector current contribution. In the inelastic 
reaction, the vector contribution vanishes for proton as well as for deuteron 
targets in the limit of $Q^2 \rightarrow 0$, and the only contribution is from 
the axial vector piece, which is only slightly affected by the deuteron 
structure. On the other hand, in the quasielastic reaction, while both vector 
and axial vector currents contribute for the nucleon case, the vector 
contribution  is completely suppressed in the deuteron. The only contribution 
left in the case of deuteron is from the axial 
vector current with an effective strength, which is strongly reduced due to 
symmetry considerations of the two nucleons in the final state\cite {14}. 
In the range of $Q^2 > 0.1$ GeV$^2$ the deuteron effects are found
to be quite small on the differential cross section $d \sigma / d q^2$ for
the inelastic reaction. The situation is then similar to the case of 
quasielastic reaction \cite{14}, where the deuteron effects are almost 
negligible in this region.

We compare the deuteron structure effects in both reactions by computing the 
ratio $R(Q^2)$ defined as

\begin{equation}
R(Q^2)=\frac{\frac{d \sigma}{d q^2}\left(\nu d \rightarrow 
\mu^- \Delta^{++} n\right)}{\frac{d \sigma}{d q^2}\left(\nu d \rightarrow 
\mu^- p p\right)}\,,
\end{equation}
and plotting it as a function of $Q^2$. In calculating $R(Q^2)$, we use the
deuteron wave function obtained from Paris potential. The differential cross
sections for the quasielastic reaction is taken from Singh and
Arenhoevel\cite{14} for the case where meson exchange currents and final
state interaction effects are neglected, in order to be consistent with our
present calculation for the inelastic reaction. In Fig. 2, we show $R(Q^2)$
for the range of low $Q^2$, where deuteron effects are known to be
important in the case of quasielastic reactions. We also show in this
figure the ratio for the equivalent reactions on the free nucleon. We see that
the ratio of the two differential cross sections on the free nucleon target
remains approximately constant for a large range of $Q^2$ considered here.
In the case of the deuteron target also the ratio $R(Q^2)$ is fairly constant
for the values of $Q^2 > 0.05$  GeV$^2$ . Note that in  the region of
$0.05 < Q^2 < 0.10$ GeV$^2$, the deuteron effects are  always less than 
$7\%$. For values of $Q^2 < 0.05$ GeV$^2$, the ratio $R(Q^2)$ increases.
This is mainly due to the decrease in the cross sections of the quasielastic
reaction. 

In the region of very low $Q^2$, the nonzero muon mass may play a role.
In order to see its effect, we have evaluated the  differential cross section
$d \sigma/d q^2$ from Eq. (13), keeping the muon mass term and the induced 
pseudoscalar form factor $C^{A}_6(Q^2)$, determined from the PCAC condition 
and given by Eq. (7). We show our results in Fig. 3 for the case of Paris
wave function. The effect of the nonzero muon mass is important in the region 
of very low $Q^2$ and is to be noticed in a fast decrease of the differential
cross section as $Q^2$ decreases and reaches a value $Q^2_{min}$, below which
the reaction is kinematically not allowed. In fact, in an earlier analysis of 
the Brookhaven experiment \cite{15}, this trend is clearly visible 
( See Fig. 11 of Ref. \cite{15} ) but, as no cross sections are quoted in this 
experiment, a direct comparison with our present theoretical results can not 
be made.

Finally, to conclude this section on the effect of deuteron structure
in the reaction  $\nu d \rightarrow \mu^- \Delta^{++} n$, we would like
to elaborate and extend the comments made by Kitagaki {\it et al} \cite{5}
about these effects and state that, at $E_{\nu}=1.6$ GeV: 

\begin{itemize}
\item The effects of deuteron structure are small for all $Q^2$, even for
$Q^2 < 0.1$ GeV$^2$, not exceeding $10\%$.
\item There is an additional reduction in the cross sections in the
region of $Q^2 \sim 0.05$ GeV$^2$ due to the nonzero muon mass, which is about 
$5\%$, and could be larger as $Q^2$ decreases further.
\end{itemize}

\subsection{Effect of the width of $\Delta$ resonance}

The analysis of Schreiner and von Hippel \cite{9} uses an S-wave 
width for the $\Delta$ resonance, which has also been used in the ANL and BNL 
experiments\cite{4,5,7}. The recent paper of HHM \cite{3}, dealing with the 
$N-\Delta$ couplings and the extraction of $C^{A}_5$, uses an expression for 
the differential cross section at $Q^2 =0$, which neglects the width of the 
$\Delta$ resonance. In this situation, it seems worthwhile to 
examine the effect of the width of the $\Delta$ resonance. Therefore, we study 
the sensitivity of the differential cross section for the process 
$\nu p \rightarrow \mu^- \Delta^{++}$ to the width of the $\Delta$ resonance 
and its energy dependence. In order to do this, we evaluate the differential 
cross section given in Eq. (8) with

\begin{itemize}
\item  P-wave $\Delta$ resonance width given in Eq. (12)

\item S-wave $\Delta$ resonance width given by \cite{9}

\begin{equation}
\Gamma = \Gamma_0 \frac{q_{cm}(W)}{q_{cm}(M')}\,,
\end{equation}

\item narrow resonance limit i.e. $\Gamma \rightarrow 0$, in which the 
differential cross section is analytically given by

\begin{equation}
\frac{d \sigma}{d q^2} = \frac{1}{64 \pi} \frac{1}{(s-M^2)^2}  
G^2 \cos^2\theta_c L_{\alpha \beta} J^{\alpha \beta}.
\end{equation}
\end{itemize}
In Fig. 4, we present the results of $R(Q^2)$ with free nucleon
target for the three cases  discussed above.
We see here that the inclusion of  the width gives a considerable reduction of
the cross section, but the detailed form of its energy dependence is not very
important when an invariant mass of $W \leq W_{cut}= 1.4$ GeV is used.
We have also found that the uncertainties in the width at the
resonance energy of about $10-15$ MeV \cite{15a} do not lead to any 
substantial change in the cross section. 

\section{ Axial Vector $N-\Delta$ couplings}

\subsection{ Neutrino Scattering Experiments}

In this section, we evaluate the value of $C^{A}_5$ using the data of
Kitagaki {\it et al.} \cite{5} on $R(Q^2)$, and use it to
describe the data of Radecky {\it et al.} \cite{7} for the differential
cross section $d \sigma/d q^2$. For this purpose, the momentum dependence of 
the vector and axial vector form factors, as given in Eqs. (4-7) has been used.

Based on the theoretical discussion presented in Sec. II on the
ratio $R(Q^2)$, we recall that  this ratio remains approximately constant in
the absence of deuteron effects. Even when the deuteron effects are included,
this ratio remains more or less constant  for $Q^2 > 0.05$ GeV$^2$. In
this region of $Q^2$, where the experimental results are available, this 
seems to be the case \cite{3,4,5}. We, therefore, assume that if the deuteron 
effects are taken out from  the quasielastic as well as from the inelastic 
reactions, this ratio remains same  even at $Q^2=0$. Of course, this limit 
is reachable only when the muon mass is neglected. Then, the cross 
sections for the quasielastic and inelastic reactions are given by \cite{16,17}

\begin{equation}
\frac{d \sigma}{d q^2}(q^2=0) = (F_A^2 + F_V^2) \frac{1}{2 \pi} G^2
\cos^2\theta_c 
\end{equation}
and
\begin{eqnarray}
\frac{d \sigma}{d q^2} (q^2=0) &=&  ({C^{A}_5})^2 \, \frac{1}{24 \pi^2}
G^2 \cos^2\theta_c
\frac{\sqrt{s} (M+M')^2 (s-M'^2)^2}{(s-M^2) M'^3} \nonumber\\[.4cm]
&\times & \int_{k'^0_{min}}^{k'^0_{max}} 
d k'^0 \frac{\Gamma(W)}{(W-M')^2+\Gamma^2(W)/4}
\end{eqnarray}
respectively; $k'^0_{min}$ and $k'^0_{max}$ are given by 

\begin{equation}
k'^0_{min}= \max\left(
\frac{s-W_{cut}^2}{2\sqrt{s}}\,,\, 0\right)\,, \;\;\;
k'^0_{max}= \frac{s-(M+m_{\pi})^2}{2\sqrt{s}}\,.
\end{equation}
Equating the ratio of these two cross sections given in 
Eqs. (17) and (18) i.e. $R(Q^2=0)$ to the extrapolated experimental 
value of $0.55 \pm 0.05$\cite{16a}, we obtain 

\begin{equation}
C^{A}_5 = 1.22 \pm 0.06
\end{equation}
Using this value of $ C^{A}_5$ and other form factors as given in 
Eqs. (4-7), we calculate the flux averaged differential cross section for the
neutrino energy spectrum of the Argonne experiment of Radecky {\it et al.}
\cite{7} and show this in Fig. 5. We see that the inclusion of deuteron and
mass effects lead to a better description of the data. It is to be
emphasized that a small reduction in the differential cross section due to
these effects is quite important in bringing out a good agreement with the
experimental data, specially in the low $q^2$ region.

In Table 1, we compare the values of these coupling constants with the
theoretical values obtained in various models. With the exception of the
quark model treatment of Liu {\it et al} \cite{2}, all the quark models
underestimate the value of $C^{A}_5$ when compared to the central values
quoted from experimental analyses. On the other hand, it is in good agreement
with the prediction of PCAC, which gives $C^{A}_5=1.15 \pm 0.01$, when the
experimental value of $g_{\Delta N \pi}=28.6 \pm 0.3$\cite{2,3} is used. It is 
expected that the various extensions of the quark models currently proposed to 
explain the quadrupole moment of $\Delta$, and the E2/M1 ratio in the photo and
electroproduction of the $\Delta$ resonance will be applied to the problem of
explaining $C^{A}_5$ and other $N- \Delta$ couplings in these models.     
   
\subsection{ Electron Scattering Experiments } 

It is possible to get information about the axial vector coupling $C^{A}_5$
from the observation of the parity violating asymmetry in the polarized 
electron scattering experiments performed in the $\Delta$ region.
The feasibility  of doing such experiments  was discussed in past by many 
authors \cite{18}, but it seems now possible to do these
experiments at the high intensity electron accelerators \cite{6,18a}. In the 
neutral current reaction $e^- + p \rightarrow e^- + \Delta^{+}$ with
polarized electron the asymmetry $A(Q^2)$ is defined as

\begin{equation}
A(Q^2) = \frac{\frac{d \sigma}{d q^2}(+1) - \frac{d \sigma}{d q^2}(-1)}
{\frac{d\sigma}{d q^2}(+1) + \frac{d \sigma}{d q^2}(-1)},
\end{equation}
where $d \sigma(\lambda)/d q^2$ is the differential cross section for an
electron with helicity $\lambda$. It has been calculated to be \cite{18a}

\begin{eqnarray}
A(Q^2) &=& - \frac{G}{2 \sqrt{2} \pi \alpha} |Q^2|  
\left[ (1 - 2 \sin^2 \theta_W) \right. \nonumber \\[.4cm]
&+& \left. (1-4 \sin^2 \theta_W) \frac{C^{A}_5}{C^{V}_3} 
\left( 1 + \frac{M'^2 + Q^2 - M^2}{2 M^2} \frac{C^{A}_4}{C^{A}_5} \right)
P(Q^2, s) \right] \\[.4cm]
&+& nonresonant \,\,\,contribution \nonumber\,,
\end{eqnarray}
where $\alpha$ is the fine structure constant and $P(Q^2,s)$ a purely 
kinematical factor. 

In principle, one can determine the value of $C^{A}_5/C^{V}_3$ from the
asymmetry measurements by selecting the kinematics where nonresonant
contributions are negligible. However, as we see from Eq. (22), the
hadronic axial vector current contribution containing $C^{A}_5$ is multiplied
by a factor $(1-4 \sin^2 \theta_W)$, which reduces the sensitivity of this
term to the asymmetry $ A(Q^2)$. This makes the extraction of $C^{A}_5$
from a measurement of the asymmetry very difficult. Even in the favourable
kinematical region of $0.5 < E_e < 1$ GeV and $Q^2 < 1.0$ GeV$^2$, this term
contributes only $(10-20)\%$, as emphasized by  Mukhopadhyay {\it et al.} 
\cite{18a}. This requires very precise measurements of $A(Q^2)$ for a 
determination of $C^{A}_5$ from parity violating asymmetry measurements.

However, there is a possibility of observing the charged current reaction
$e^- +p \rightarrow  \Delta^0 +\nu $  with unpolarized electrons
through the detection of the protons and pions from the decay of the $\Delta$
resonance \cite{19}. At the incident electron energy of 4 GeV, the differential
cross section $d \sigma/d q^2$
in the forward direction near $Q^2 = 0$ is estimated to be
$2.\,10^{-39}$ cm$^2$/GeV$^2$. For  an incident intensity of about
$ 2.\,10^{38}$cm$^2$/sec \cite{18a} and $Q^2$ bin of $0.05$ GeV$^2$, one would 
expect 72 events  per hour for the production of $\Delta^0$, assuming $100\%$
efficiency of the detector. One third of these $\Delta 's$ will produce
negatively charged pions and protons, which can be easily observed. Since in 
the region of $Q^2 \sim 0$, $C^{A}_5$ gives the dominant contribution, its 
determination from the weak charged current experiment of $\Delta $ production 
seems feasible.

\subsection{ Photo and Electro-pion production Experiments}

It is well known that, in the threshold region of photo  and electro pion
production from the nucleon, the matrix element of these processes in the soft
pion limit is related with the nucleonic matrix element of the axial vector
current using the methods of current algebra and the PCAC. This relation
has been exploited to obtain information about the axial vector form factor
of the nucleon \cite{21}. In a similar way, threshold pion
production in the processes $ e^- + p \rightarrow e^- + \Delta^+ + \pi^0$ and
$\gamma +p \rightarrow \Delta^{++} +\pi^-$ is related, in the soft pion limit,
with the $N-\Delta$ transition matrix element of the axial vector
current. The axial vector transition form factors can, in principle, be
determined from these processes in the limit of soft pions. Such attempts
have been made in past and they yield $C^{A}_5 = 1.1 \pm 0.2$\cite {21a}   

However, in this case, the treatment of higher resonances and their effective 
couplings used for evaluating the matrix elements of the time ordered
product of the vector and axial vector current operators occurring in the 
LSZ reduction involve many approximations, which need further justification. 
Recently, there has been some progress in calculating the 
contribution of higher resonances to the production of two pions in the photo 
and electroproduction processes using effective Lagrangians \cite {22}. It 
should be possible to isolate the dominant contributions from higher 
order resonances, which are relevant for the $\Delta \pi$ production in the 
soft pion limit. This will help to reduce the theoretical uncertainties in 
the application of the methods of PCAC and current algebra to the processes 
where a $\Delta$ resonance is produced. In addition, when dealing with the 
$\Delta $ resonance, its width has to be properly taken into account as 
remarked by Bartl {\it et al.} \cite {21a}, and also shown by us in the weak 
charged current production of the $\Delta $ resonance. The  analysis of  Bartl 
{\it et al.} \cite {21a} uses the older data which suffers from poor 
statistics. When the results of a recent experiment proposed at TJNAF
\cite {23} become available in near future, it will be possible to get precise 
information about the axial vector coupling $C^{A}_5$ and its momentum 
dependence.  

\section{Summary and outlook}

We have calculated the effect of deuteron structure and width of the $\Delta$
resonance in the differential cross section for the reaction
$\nu d \rightarrow \mu^- \Delta^{++} n$ and find that these effects are
small, but important in order to explain the experimental results at low $q^2$,
where they were initially expected to be
important. Furthermore, in the region of very low $q^2$, the muon mass, which
is usually neglected in the calculations, also reduces the cross section.

The effect of the width of the $\Delta$ resonance on the cross section is
important and plays a crucial role in bringing out good agreement with the 
experimental data. The detailed shape and 10-15\% uncertainty in the width of 
the resonance does not affect the cross sections very much. 

The axial vector $N-\Delta$ coupling $C^{A}_5$ is extracted from the BNL
data on $\nu d \rightarrow \mu^- \Delta^{++} n$, incorporating the
effect of the deuteron structure and the width of $\Delta$ resonance. 
This value of $C^{A}_5$ is found to be larger than the values predicted in 
most of the quark models and is consistent with the prediction of PCAC and
Adler's model.

Finally, we have discussed the possibility of determining this coupling from 
electron scattering experiments, and find that electroproduction and weak 
charged current of $\Delta$ resonance are more suitable than asymmetry 
measurements in the polarized electroproduction of $\Delta$.

\acknowledgments

This work has been partially supported by DGYCIT contract No.
PB 96-0753. One of us (L.A.R.) acknowledges financial support from the
Generalitat Valenciana. S.K.S. has the pleasure of thanking Prof. Oset for
his hospitality at the University of Valencia, and acknowledges financial
support from the Ministerio de Educaci\'on y Cultura of Spain in his
sabbatical stay.

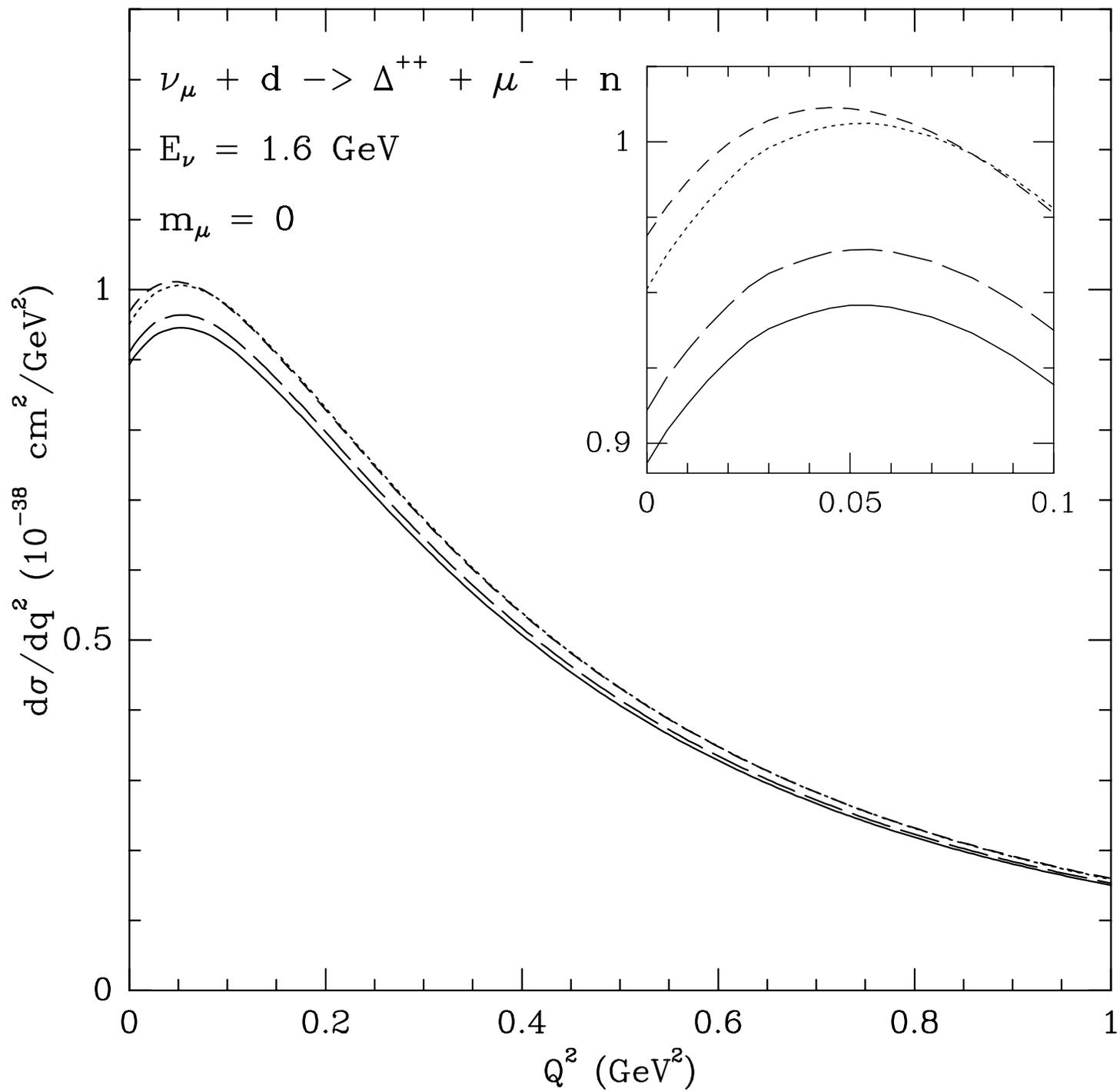
\begin{figure}
\caption{Differential cross section for weak charged current neutrino
production of $\Delta$ on deuteron. In the short-dashed line, deuteron effects 
are neglected while dotted, long-dashed and solid lines include these
effect using Hulthen, Bonn and Paris deuteron wave functions respectively.} 
\end{figure}

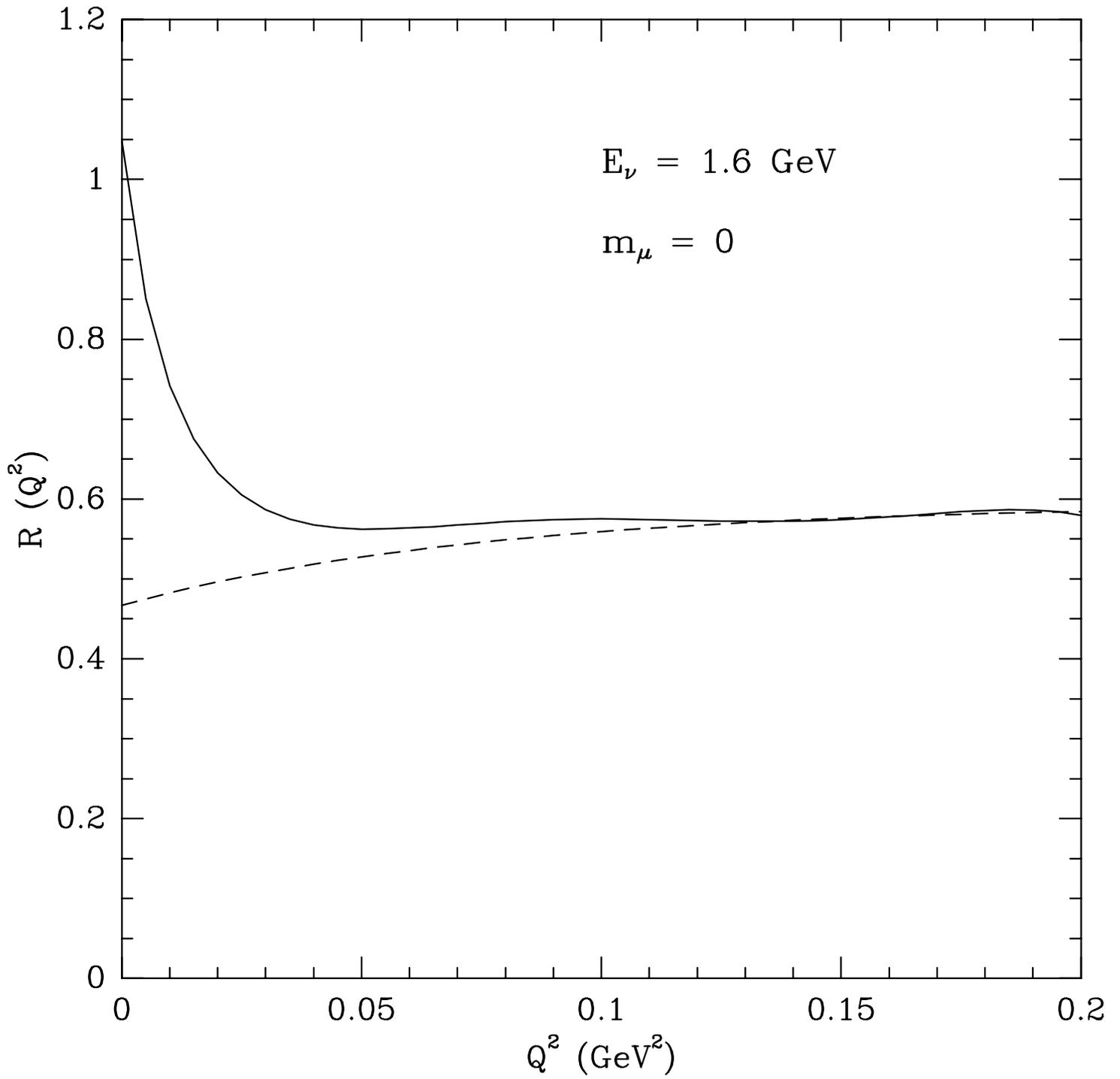
\begin{figure}
\caption{Ratio of $\Delta$ production and quasielastic reactions differential 
cross sections with (solid line) and without (dashed line) deuteron effects.} 
\end{figure}

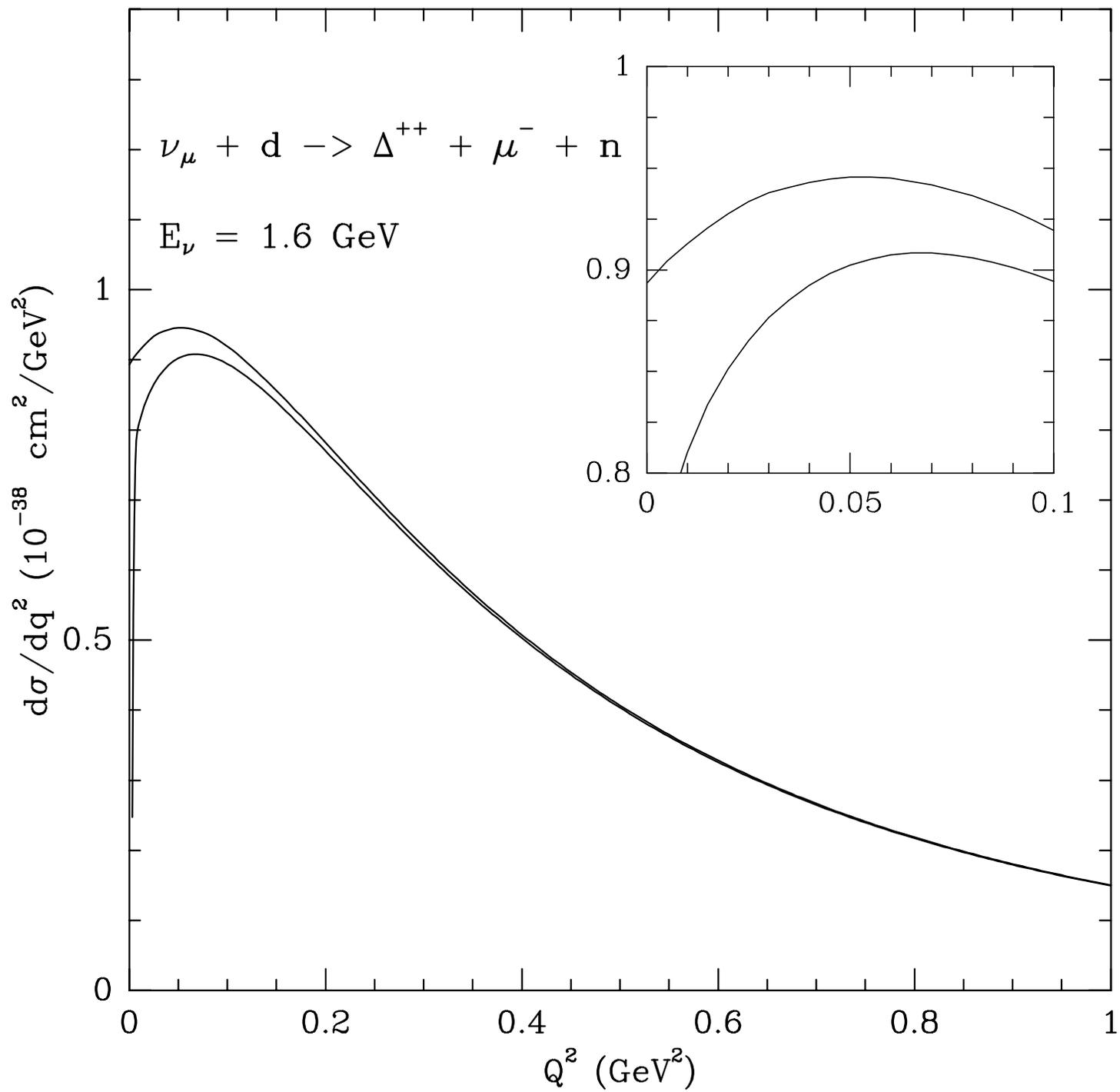
\begin{figure}
\caption{Effect of the muon mass on the differential cross section for the 
$\nu d\rightarrow \mu^- \Delta^{++}n$ reaction. In the upper line muon mass
is neglected while it is considered in the lower one. Both curves include
deuteron effects using the Paris parametrization of deuteron wave function.} 
\end{figure}

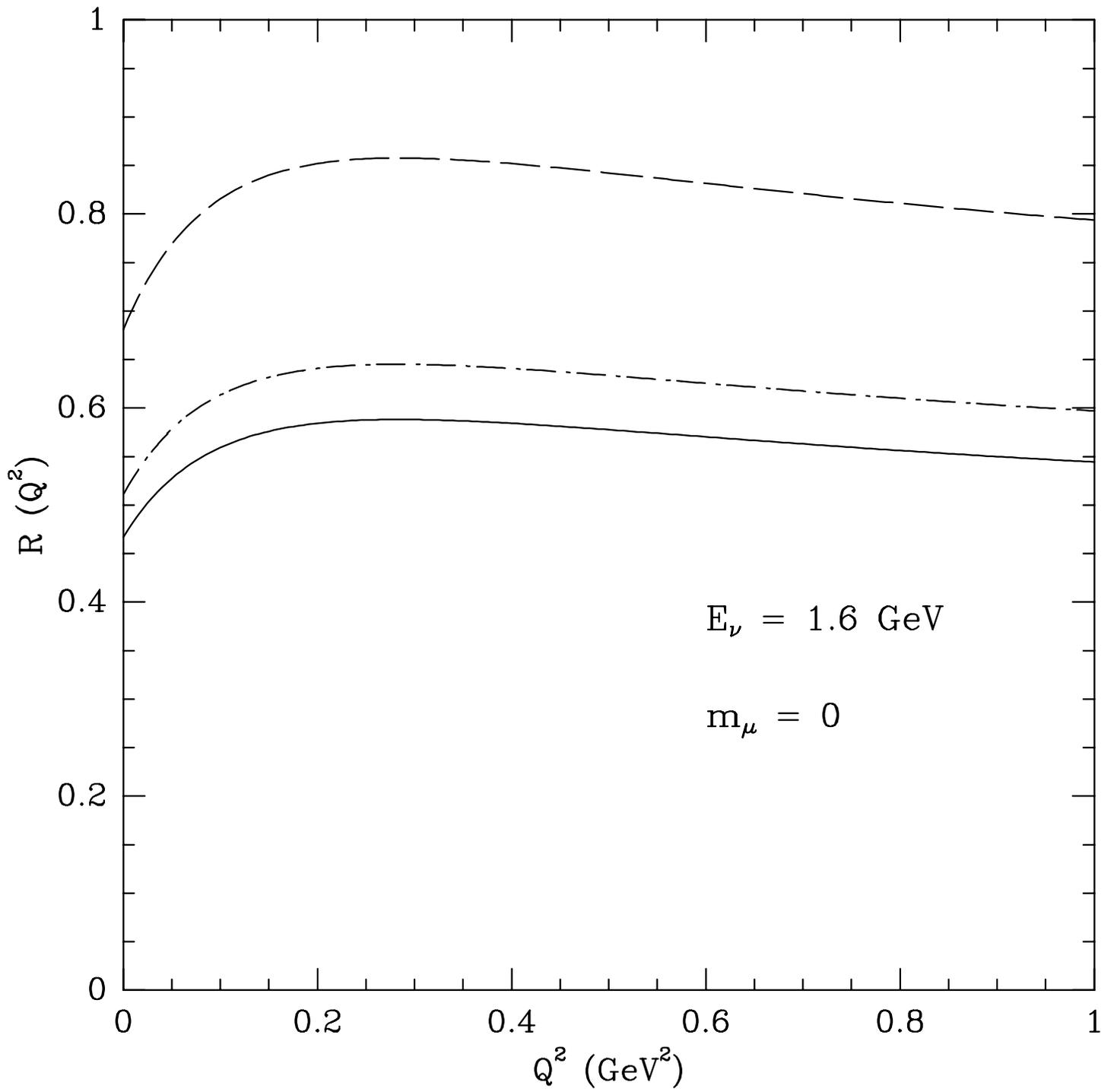
\begin{figure}
\caption{Effect of $\Delta$ width in $R(Q^2$): the solid line corresponds to
a P-wave width, the dash-dotted line, to an S-wave width and the dashed line, 
to the case of zero width resonance. Deuteron effects have been neglected in 
all curves.} 
\end{figure}

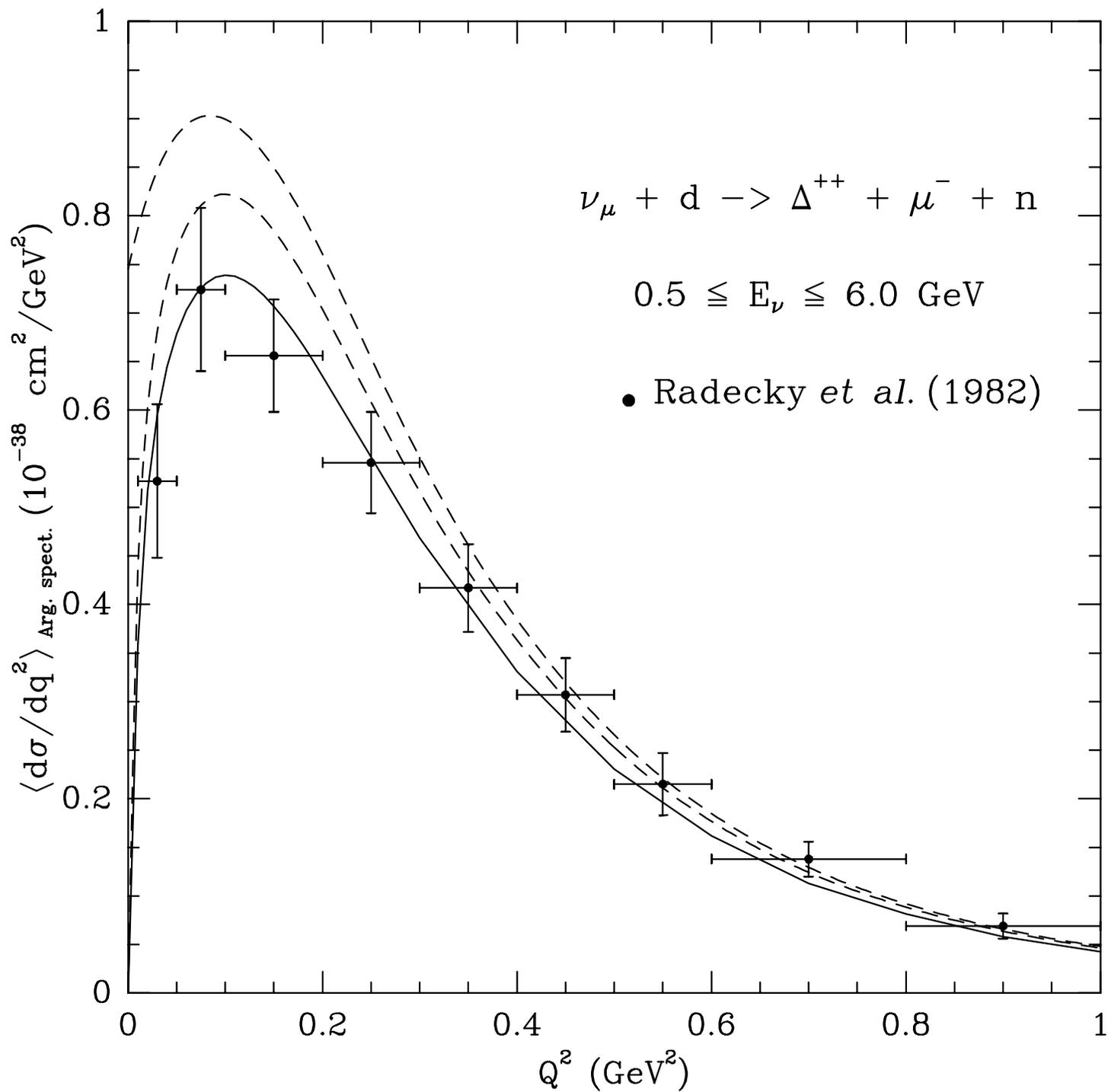
\begin{figure}
\caption{Differential cross section for weak charged current neutrino
production of $\Delta$ on deuteron, averaged over the spectrum of ANL
experiment, compared to the experimental results given in 
Ref.\protect\cite{7}. The solid curve includes both nonzero muon mass and
deuteron effects. Upper dashed curve neglects muon mass and deuteron effects.
Lower dashed curve neglects only deuteron effects.} 
\end{figure}

\mediumtext
\begin{table}
\caption{The numerical values of axial $N-\Delta$ coupling $C_5^A$ in
various quark model and empirical approaches. The earlier, prior to 1973,
evaluations of these couplings in these approaches have been summarized by
Schreiner and von Hippel\protect\cite{9} and Llewellyn Smith\protect\cite{8}.} 
\begin{tabular}{cl}
&$C_5^A$ \\
\tableline
Quark Model approaches& 0.97\protect\cite{24,25}, 0.83\protect\cite{26},
1.17\protect\cite{2}, 1.06\protect\cite{27}, 0.87 \protect\cite{3}\\
Empirical approaches& 1.15$\pm$0.23\protect\cite{4}, 
1.39$\pm$ 0.14\protect\cite{3}, 1.1$\pm$ 0.2\protect\cite{21a}, 
1.22$\pm$ 0.06\tablenotemark[1] \\
\end{tabular}
\tablenotetext[1]{Present result.}
\end{table}


\begin{references}
\bibitem[*]{singhper} Permanent Address: Physics Department, Aligarh
Muslim University, Aligarh, India 202002.

\bibitem{1} R. M. Davidson, N. C. Mukhopadhyay and R. S. Wittman, Phys. Rev.
{\bf D43}, 71 (1991); Phys. Rev. Lett. {\bf 56}, 804 (1986); R. M. Davidson
and N. C. Mukhopadhyay, Phys. Rev. {\bf D42}, 20 (1990); S. Capstick and G.
Karl, Phys. Rev. {\bf D41}, 2767 (1990); T. D. Cohen and W. Bronioski,
Phys. Rev {\bf D34}, 3472 (1986); G. Kalberman and J. M. Eisenberg Phys.
Rev. {\bf D28}, 71 (1983); N. Isgur, G. Karl and R. Koniuk, Phys. Rev. {\bf
D25}, 2394 (1982) and references therein.      
\bibitem{2} For a recent overview, see  J. Liu, N. C. Mukhopadhyay and 
L. Zhang, Phys. Rev. {\bf C52}, 1630 (1995). 
\bibitem{3} T. R. Hemmert, B. R. Holstein and N. C. Mukhopadhyay, Phys.
Rev. {\bf D51}, 158 (1995). 
\bibitem{4} S. J. Barish {\it et al.}, Phys. Rev. {\bf D19}, 2521 (1979).
\bibitem{5} T. Kitagaki {\it et al.}, Phys. Rev. {\bf D42}, 1331 (1990). 
\bibitem{6} V. D. Burkert, in {\it Perspectives in the Structure of Hadronic
Systems}, edited by M. N. Harakeh {\it et al.} (Plenum Press, New York,
1994),p. 101; F. E. Maas {\it et al.}, in {\it Proceedings of the Erice
Summer School on the Spin structure of the Nucleon} ( Erice, Sicily, 1996); 
M. J. Musolf {\it et al.}, Phys. Rep. {\bf 239}, 1 (1994); 
{\it Proceedings of the Workshop on Parity Violation in electron
scattering}, edited by E. J. Beise and R. Mckeown ( World Scientific,
Singapore, 1991).
\bibitem{7} G. M. Radecky {\it et al.}, Phys. Rev. {\bf D25}, 1161 (1982).
\bibitem{8} C. H. Llewellyn Smith, Phys. Rep. {\bf 3}, 261 (1971) and
references therein.
\bibitem{9a} S. K. Singh, M. J. Vicente Vacas and E. Oset, Phys. Lett. {\bf
B416}, 23 (1998); G. L. Fogli and G. Nardulli, Nucl. Phys. {\bf B160}, 116 
(1979); S. L. Adler, Phys. Rev. {\bf D12}, 2644 (1975); S. L. Adler, 
Ann. Phys. (N.Y.) {\bf 50}, 189 (1968).
\bibitem{9} P. A. Schreiner and F. von Hippel, Nucl. Phys. {\bf B58}, 333 
(1973). 
\bibitem{10} P. Allen {\it et al.}, Nucl. Phys. {\bf B176}, 269 (1980);
J. Bell {\it et al.}, Phys. Rev. Lett. {\bf 41}, 1008 (1978 ); {\bf 41},
1012 (1978); P. Zucker, Phys. Rev. {\bf D4}, 3350 (1971); 
J. Bijtebier, Nucl. Phys {\bf B21}, 158 (1970).
\bibitem{33} E. Oset, H. Toki and W. Weise, Phys. Rep. {\bf 83}, 281 (1982).
\bibitem{15a} R. H. Barnett {\it et al.}, Phys. Rev. {\bf D54}, 1 (1996).
\bibitem{14} S. K. Singh and H. Arenhoevel, Z. Phys. {\bf A324}, 347 (1986);
S. L. Mintz, Phys. Rev. {\bf D13}, 637 (1976); R. Tarrach and P. Pascual,
Nuovo Cimento {\bf 18A}, 760 (1973); J. Bernabeu and P. Pascual, Nuovo Cimento 
{\bf 10A}, 61 (1972); S. K. Singh, Nucl. Phys. {\bf B36} 419 (1971).
\bibitem{11} L. Hulthen and M. Sugawara, Handbuch der Physik ( Springer 
Verlag, Berlin, 1957), vol. 39.
\bibitem{13} M. Lacombe {\it et al.}, Phys. Lett. {\bf 101B}, 139 (1981).
\bibitem{12} R. Machleidt, K. Holinde and C. Elster, Phys. Rep. 
{\bf 149}, 1 (1987). 
\bibitem{15} T. Kitagaki {\it et al.}, Phys. Rev. {\bf D34}, 2554 (1986).
\bibitem{16} Our expression in Eq. (18), when taken in the zero width 
limit, is at variance with the results of \cite{3}, but is in agreement with 
the result of Barish {\it et al.}\cite{4}. This expression also agrees with 
the results of Albright and Liu\cite{17} (see Eq. (3.15)), 
when the Eq. (2.12) is used along with the conversion table 
of the various definitions for the transition form factors given by 
Llewellyn Smith\cite{8}.
\bibitem{17} C. H. Albright and L. S. Liu, Phys. Rev. {\bf 140} ,B78 (1965).
\bibitem{16a} Here we use the value quoted by Kitagaki{\it et al.}\cite{5},
with error estimated from their data on $R(Q^2)$.  
\bibitem{24} F. Ravndal, Nuovo Cimento {\bf A18}, 385 (1973).
\bibitem{25} J. G. Korner, T. Kobayashi, and C. Avilez, Phys. Rev. {\bf
D18}, 3178 (1972).
\bibitem{26} A. Le Yaouanc {\it et al.}, Phys. Rev. {\bf D15}, 2447 (1977).
\bibitem{27} M. Beyer Habilitation Dissertation ( University of Rostock, 
Germany, 1997), p. 60. 
\bibitem{18} L. M. Nath, K. Schilcher and  M. Kretzschmar, Phys. Rev.
{\bf D25}, 2300 (1982); D. R. T. Jones and S. T. Petcov, Phys. Lett. {\bf
91B}, 137 (1981).
\bibitem{18a} N. C. Mukhopadhyay {\it et al.}, nucl-th/9801025;
H. W. Hammer and D. Drechsel, Z. Phys. {\bf A353}, 321 (1995).
\bibitem{19} L. Alvarez-Ruso, S. K. Singh and M. Vicente-Vacas,  Phys.Rev.
{\bf C}, in print. 
\bibitem{21} E. Amaldi, S. Fubini and G. Furlan, Pion electroproduction,
Springer Tracts in Modern Physics (Springer, Berlin, 1977), vol. 83.  
\bibitem{21a} A. Bartl, K. Wittman, N. Paver and C. Verzegnassi, Nuovo Cimento
{\bf 45A}, 457 (1978); A. Bartl, N. Paver, C. Verzegnassi and S. Petrarca, 
Lett. Nuovo Cimento {\bf 18}, 588 (1977).  
\bibitem{22} J. C. Nacher and E. Oset, nucl-th/9804006; 
J. A. Gomez Tejedor and E. Oset, Nucl. Phys. {\bf A571}, 400 (1994).
\bibitem{23} L. Elouadrhiri {\it et al.}, CEBAF Experiment E-94-005.
\end{references}
\end{document}